\pdfoutput=1

\documentclass[prl,twocolumn]{revtex4-1}

\usepackage{amsmath}
\usepackage{amsfonts}
\usepackage{array}
\usepackage{graphicx}
\usepackage{lineno}
\usepackage[table]{xcolor}
\newcolumntype{C}[1]{>{\centering\arraybackslash$}m{#1 in}<{$}}

\begin{document}

\title{Symmetry Configuration Mapping for Representing Quantum Systems on Quantum Computers}

\author{Sean A. Fischer}
\author{Daniel Gunlycke}
\email{daniel.gunlycke@nrl.navy.mil}
\affiliation{U.S. Naval Research Laboratory, Washington, DC 20375, USA}

\begin{abstract}
Quantum computing has the potential to significantly speed up complex computational tasks, and arguably the most promising application area for near-term quantum computers is the simulation of quantum mechanics. To make the most of our limited quantum computing resources, we need new and more compact algorithms and mappings. Whereas previous work, including recent demonstrations, has focused primarily on general mappings for fermionic systems, we propose instead to construct customized mappings tailored to the considered quantum mechanical systems. Specifically, we take advantage of existing symmetry, which we build into the mappings a priori to obtain optimal compactness. To demonstrate this approach, we have performed quantum computing calculations of the fluorine molecule, in which we have mapped 16 active spin-orbitals to 4 qubits. This is a four-fold reduction in the qubit requirement, as compared to the standard general mappings. Moreover, our compact system-to-qubits mappings are robust against noise that breaks symmetry, thereby reducing non-statistical errors in the computations. Furthermore, many systems, including F$_{2}$, are described by real Hamiltonians, allowing us to also reduce the number of single-qubit operations in the hardware-efficient ansatz for the quantum variational eigensolver by roughly a factor of three.
\end{abstract}

\maketitle

To understand the behavior of an isolated, non-relativistic system, we solve a quantum mechanical equation of motion such as the Schr\"odinger equation
\begin{equation}
i\hbar\frac{\partial}{\partial t}|\Psi(t)\rangle=\hat H|\Psi(t)\rangle,
\label{eq:SE}
\end{equation}
where $\hat H$ is the Hamiltonian describing the system and $|\Psi(t)\rangle$ is the quantum state at time $t$.  While finding the exact solution is with few exceptions computationally hard on classical computers, Eq.\,(\ref{eq:SE}) can be solved efficiently on quantum computers.\cite{Benioff80,Feynman82,Lloyd96_1073,Abrams97}

Current research directed towards using quantum computers to solve quantum mechanical problems has mostly focused on a second quantized description that use an occupation number representation to describe quantum states\cite{Abrams97} and the Jordan-Wigner (JW) transformation\cite{Jordan28} to represent the dynamics of the system on a quantum computer.\cite{Ortiz01,Kitaev02_210,White10_106,AspuruGuzik11_735,Love12_224109,OBrien14_4213,Whitfield17_032332,Temme17_08213,Wehner18_063010}  While mapping the Fock space onto the Hilbert space of a quantum register comprised of qubits has the advantage of known and efficient transformations, it is not the most efficient mapping in terms of the number of qubits needed for many problems of interest.\cite{HeadGordon05_1704} Efficient mappings are particularly important for the current generation of quantum computers that demand the efficient use of limited quantum resources.

Herein, we apply the symmetry configuration mapping (SCM) method,\cite{NRL_SCM} which begins by constructing a complete set of symmetry-adapted many-particle basis states $\{|\Phi^\Sigma_\ell\rangle\}$, which we refer to as symmetry configuration basis states, where $\Sigma$ denotes symmetry configurations and $\ell$ enumerates symmetry-equivalent basis states.  In the symmetry configuration basis, the Hamiltonian of the system can be expressed as the direct sum
\begin{equation}
\hat H=\bigoplus_{\Sigma}\hat H^{\Sigma},
\end{equation}
where $\hat{H}^{\Sigma}$ are symmetry configuration Hamiltonians of rank $\Lambda^{\Sigma}$.  This decomposition allows us to solve the eigenvalue equation
\begin{equation}
\hat H^{\Sigma}|\Psi^\Sigma_\lambda\rangle=E^\Sigma_\lambda|\Psi^\Sigma_\lambda\rangle
\label{eq:EE}
\end{equation}
independently for each symmetry configuration $\Sigma$, where $|\Psi^\Sigma_\lambda\rangle$ are eigenstates with energies $E^\Sigma_\lambda$.  Moreover, we often need to consider only a subset of symmetry configurations that are known a priori from, e.g., spectroscopic data or classical calculations.  For ground-state properties in particular of a system with known symmetry, we only need to solve Eq.\,(\ref{eq:EE}) for a single symmetry configuration.

Using the symmetry configuration basis in the SCM method outlined above, we can represent the symmetry configuration Hamiltonians on the Hilbert space of the quantum register as linear combinations of tensor products of Pauli operators:
\begin{equation}
\hat H^\Sigma=\sum_{(r_q)}c_{(r_q)}\hat\sigma_{r_1}\otimes\hat\sigma_{r_2}\otimes\ldots\otimes\hat\sigma_{r_Q},
\label{eq:qham}
\end{equation}
where $(r_q)$ are indexed families over all qubits $q\in\{1,2,...,\mathrm Q^\Sigma\}$ with each $r_q\in\{0,x,y,z\}$, $\hat\sigma_0$ is the identity operator, and $\hat\sigma_x$, $\hat\sigma_y$, $\hat\sigma_z$ are Pauli operators. For any given $\Sigma$, the number of qubits required to map our symmetry configuration basis states $|\Phi^\Sigma_\ell\rangle$ to the basis states of our quantum register is
\begin{equation}
\mathrm Q^\Sigma=\lceil\log_2\Lambda^\Sigma\rceil,
\label{eq:qubits}
\end{equation}
which can be substantially less than in other mappings.  See Table~\ref{tab:resources}. On this form, solutions to Eq.\,(\ref{eq:EE}) can be obtained using a quantum computer.  Various algorithms have been proposed,\cite{Yuan18_10402} including the quantum variational eigensolver,\cite{OBrien14_4213} which essentially applies the variational method of quantum mechanics to quantum computing to obtain ground-state properties.

\begin{figure}
   \centering
   \includegraphics[width=8.5cm]{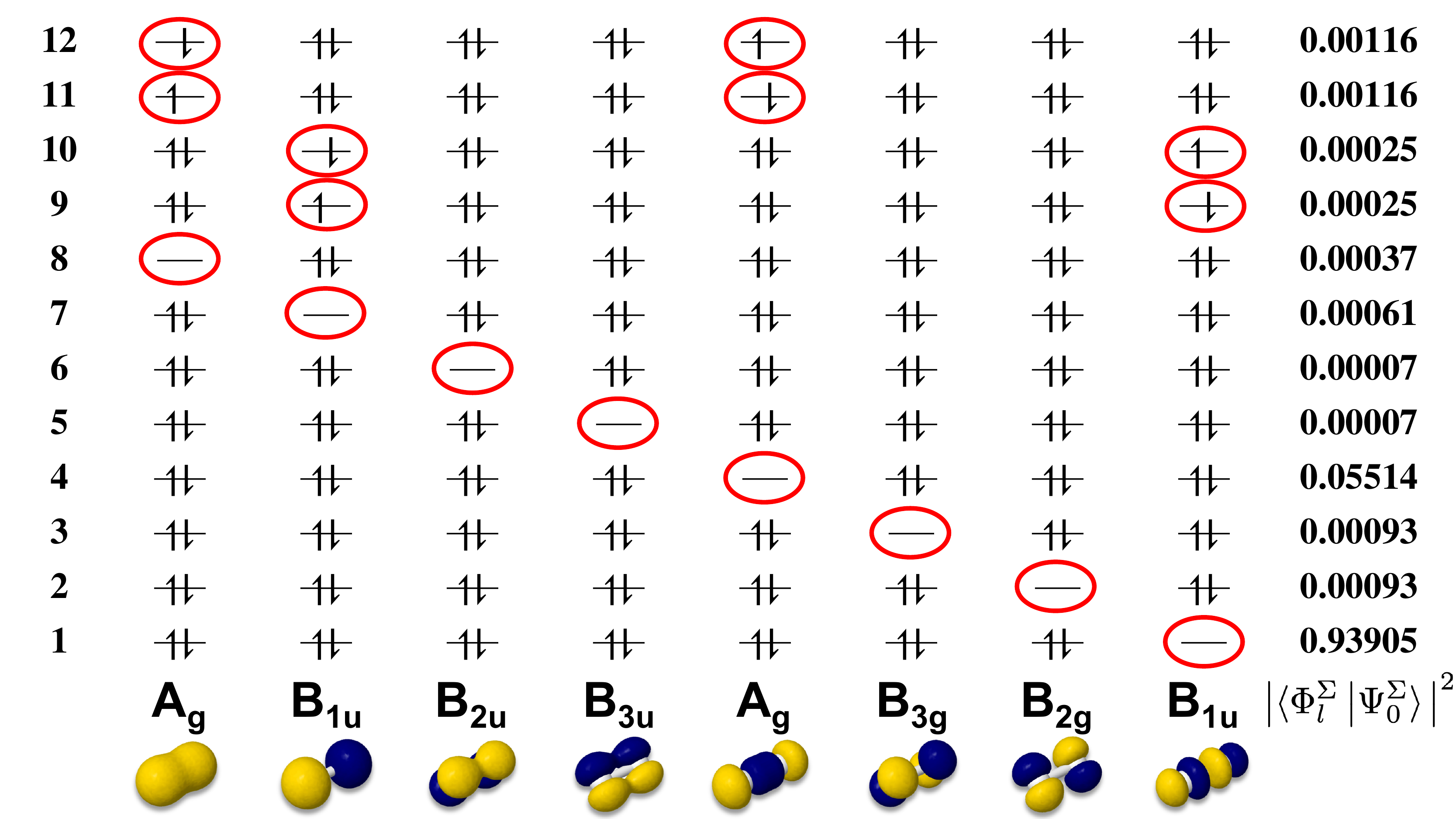}
   \caption{The 12 symmetry configuration basis states in the symmetry configuration $\Sigma=(14,0,\mathrm{A_g})$. The valence molecular orbitals of F$_2$ and their irreducible representations are given at the bottom. The numbers in the last column indicate the modulus squared of the overlap of each symmetry configuration basis state with the exact F$_2$ ground state at the equilibrium geometry.}
   \label{fig:MO}
\end{figure}

\begin{table}
   \centering
   \begin{tabular}{C{1.2}|C{1.2}|C{0.6}}
   \hline
   $Constraints$ & $Configurations$ & $Qubits$ \\
   \hline \hline
   $None$ & 65536 & 16 \\
   ${N}$ & 120 & 7 \\
   ${N, S$_{z}$}$ & 64 & 6 \\
   \rowcolor{gray!30}${N, S$_{z}$, $\Gamma$}$ & 12 & 4 \\
   ${N, S, S$_{z}$, $\Gamma$}$ & 10 & 4 \\
   \hline
   \end{tabular}
   \caption{Number of configurations and qubits needed for F$_{2}$ when various constraints are applied. With a frozen core in the minimal atomic basis, F$_{2}$ has 8 spatial orbitals, 16 spin-orbitals, and 14 electrons (N). The top row represents the standard approach while the remaining rows represent variants of SCM. We consider the cases when the spin-projection S$_{z}$ is fixed to 0 and when the spatial symmetry $\Gamma$ is fixed to the A$_{g}$ irreducible representation. The highlighted row is the setup used in the present study.}
   \label{tab:resources}
\end{table}
Herein, we demonstrate the SCM method by solving for the ground state of the F$_{2}$ molecule. F$_{2}$ has 18 electrons and a minimal basis consisting of 20 spin-orbitals; however, we have adopted a frozen-core approximation to eliminate 4 electrons and 4 spin-orbitals from the configuration interaction active space (Fig.~\ref{fig:MO}). In general, we can construct the symmetry configuration basis states $|\Phi^\Sigma_\ell\rangle$ such that: (1) they are simultaneous eigenstates of the total number operator $\hat N$ and the total spin operators $\hat S^2$ and $\hat S_z$ with quantum numbers $\mathrm N$, $\mathrm S$, and $\mathrm S_z$, respectively, and (2) they form a complete basis for the irreducible representation $\Gamma$ of the spatial symmetry group of the system.  This lets us express our symmetry configurations as $\Sigma=(\mathrm N, \mathrm S, \mathrm S_z, \Gamma)$. In our F$_{2}$ demonstration however, we exclude total spin $\mathrm S$ and focus on configurations on the form $\Sigma=(\mathrm N, \mathrm S_z, \Gamma)$. Even forgoing $S$, we have still achieved a four-fold reduction in the required number of qubits, as shown in Table~\ref{tab:resources}.
 
There are only twelve symmetry configuration basis states $|\Phi^{(\mathrm N, \mathrm S_z, \Gamma)}_\ell\rangle$ for the symmetry configuration of the lowest-energy state, which has S$_{z} = 0$ and the irreducible representation $\mathrm{A_{g}}$ of the $\mathrm{D_{2h}}$ point group. These symmetry configuration basis states $|\Phi^{(14,0,\mathrm{A_g})}_\ell\rangle$, illustrated in Fig.~\ref{fig:MO}, can be understood by focusing on hole states describing unoccupied orbitals. For a basis state to have the appropriate A$_{g}$ symmetry, the two unoccupied orbitals must belong to the same irreducible representation, which is only satisfied by the twelve states in Fig.~\ref{fig:MO}.  Equation\,(\ref{eq:qubits}) tells us that we only need 4 qubits to map these 12 symmetry configuration basis states, instead of the 16 qubits required to map all the 65,536 basis states of the Fock space.

The factor of 4 savings in the number of qubits could also lead to a savings in the number of variation parameters needed. For the hardware-efficient ansatz,\cite{Gambetta17_242} the number of parameters scales linearly with the number of qubits. This results in the same relative savings in the number of parameters, at a given algorithm depth, as the number of qubits. Furthermore, it is possible that the more compact representation generated by the SCM method could require a shallower algorithm to reach the desired accuracy.

For the F$_2$ calculations, we used the hardware-efficient algorithm of Kandala et al., which consists of entangling two-qubit operations sandwiched between layers of single qubit operations to prepare our trial states.\cite{Gambetta17_242} Specifically, we used the CNOT gate to entangle the qubits. Kandala et al. used arbitrary single-qubit operations in their implementation of the hardware-efficient ansatz $U^{q,i}(\boldsymbol{\theta}) = R_{z}(\theta^{q,i}_{1})R_{x}(\theta^{q,i}_{2})R_{z}(\theta^{q,i}_{3})$, where $q$ is again a qubit index and $i$ identifies the depth layer of the quantum circuit. With the qubits all initialized in their ground states, the first rotation around the $z$-axis is inconsequential, and thus, the number of optimization parameters is $\mathrm D^\Sigma=(3d+2)\mathrm Q^\Sigma$, where $d$ is the depth of the circuit.  Not all of these parameters however, are generally needed as we do not seek arbitrary solutions, but rather particular solutions.

\begin{figure}
   \centering
   \includegraphics[width=8.5cm]{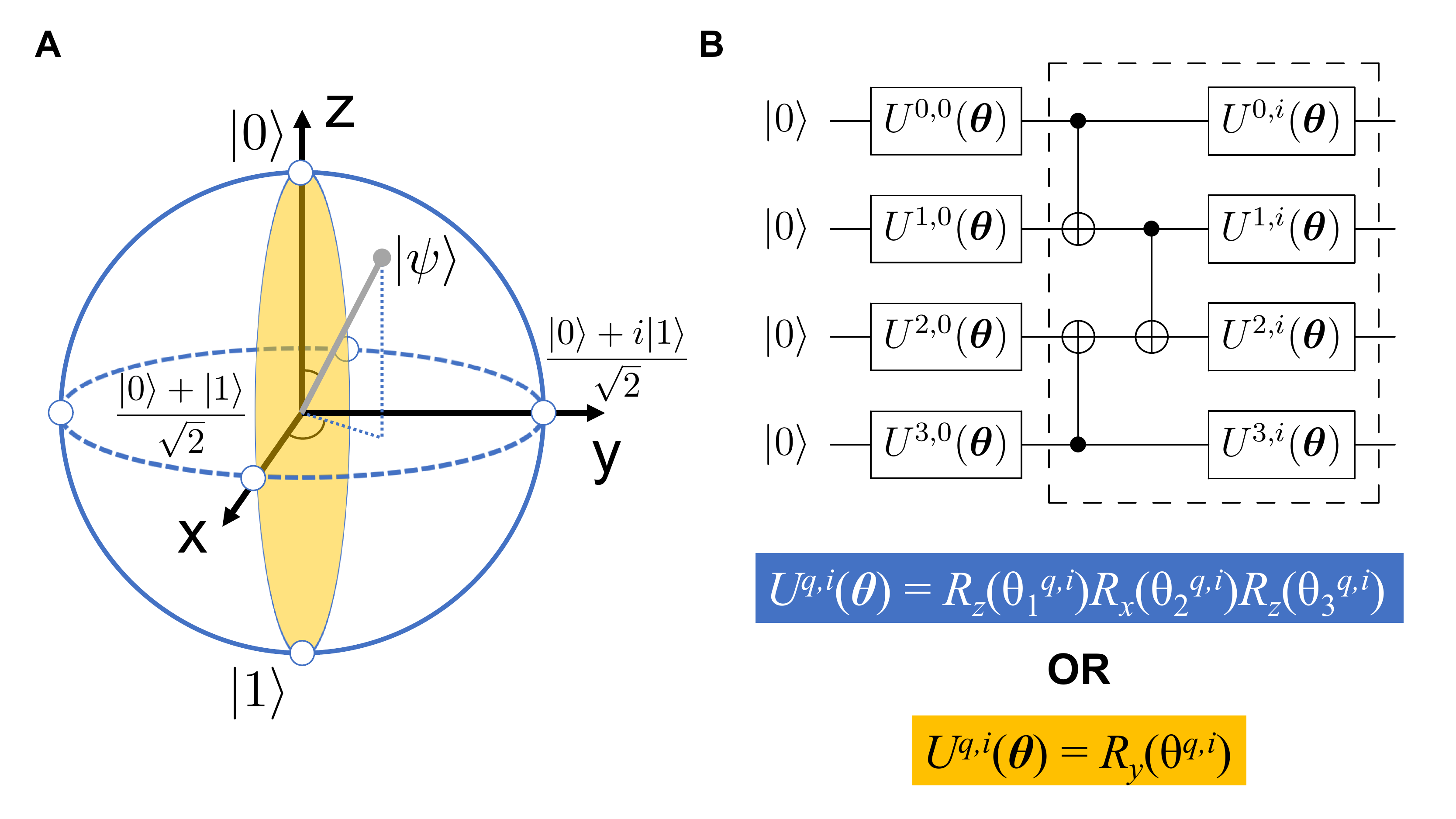}
   \caption{(A) Bloch sphere representation of a single-qubit state. The state remains real as long as it is confined to the \textit{xz}-plane, allowing us to restrict our single-qubit rotations to those around the \textit{y}-axis. (B) The hardware-efficient circuit ans\"atze considered in the present work. The rotations in blue produce complex trial states, while the rotations in gold produce only real trial states.}
   \label{fig:BS}
\end{figure}
For many systems of interest in quantum mechanics, the Hamiltonian contains only real terms. If we have a real Hamiltonian, then the decomposition of the matrix representation of the Hamiltonian can be achieved with real eigenvectors.\cite{Wilf81_49} Because our entangling gate operator has a real representation, we can without loss of generality restrict the single-qubits operations to those that leave each qubit state in the real subspace.\cite{Moll18} Visually, we can see in the Bloch sphere representation in Fig.~\ref{fig:BS}A that a qubit state remains real as long as it is confined to the $xz$-plane. This means that we can restrict our single-qubit operations to $U^{q,i}(\boldsymbol{\theta}) = R_{y}(\theta^{q,i})$.  The number of optimization parameters, and needed single-qubit gates, is then
\begin{equation}
\mathrm D^\Sigma=(d+1)\mathrm Q^\Sigma,
\label{eq:parameters}
\end{equation} 
which is nearly a reduction by a factor of 3, in addition to the reduction by a factor of 4 in $\mathrm Q^\Sigma$, and the fact that $d$ generally decreases with decreasing $\mathrm Q^\Sigma$.

Through simulation of our quantum circuits, we found no statistically significant differences in the calculated energies or needed algorithm depths, when comparing the arbitrary single-qubit rotations with those that enforce the real-space restriction. Additionally, we found that an algorithm depth of 1 was sufficient to reach an average accuracy of 1.07 kcal/mol across the potential energy curve of F$_{2}$, with a maximum error of 3.03 kcal/mol. A circuit depth of 2 resulted in the entire curve being calculated to within chemical accuracy (maximum error less than 1 kcal/mol).

Having demonstrated that chemical accuracy with short circuit depth can be achieved with our setup, we now turn to analyze the performance on an actual quantum computer. To this end, we have run experiments on IBM's Poughkeepsie quantum computer. We used an algorithm depth of 1 in order to minimize decoherence errors and reduce the buildup of gate errors. To account for measurement errors in our results, we constructed a calibration matrix whose columns correspond to the probability to measure each computational basis state for the preparation of a given computational basis state. This matrix was then inverted and used to correct the measured probabilities for the hardware-efficient ansatz.

\begin{figure}
   \centering
   \includegraphics[width=8.5cm]{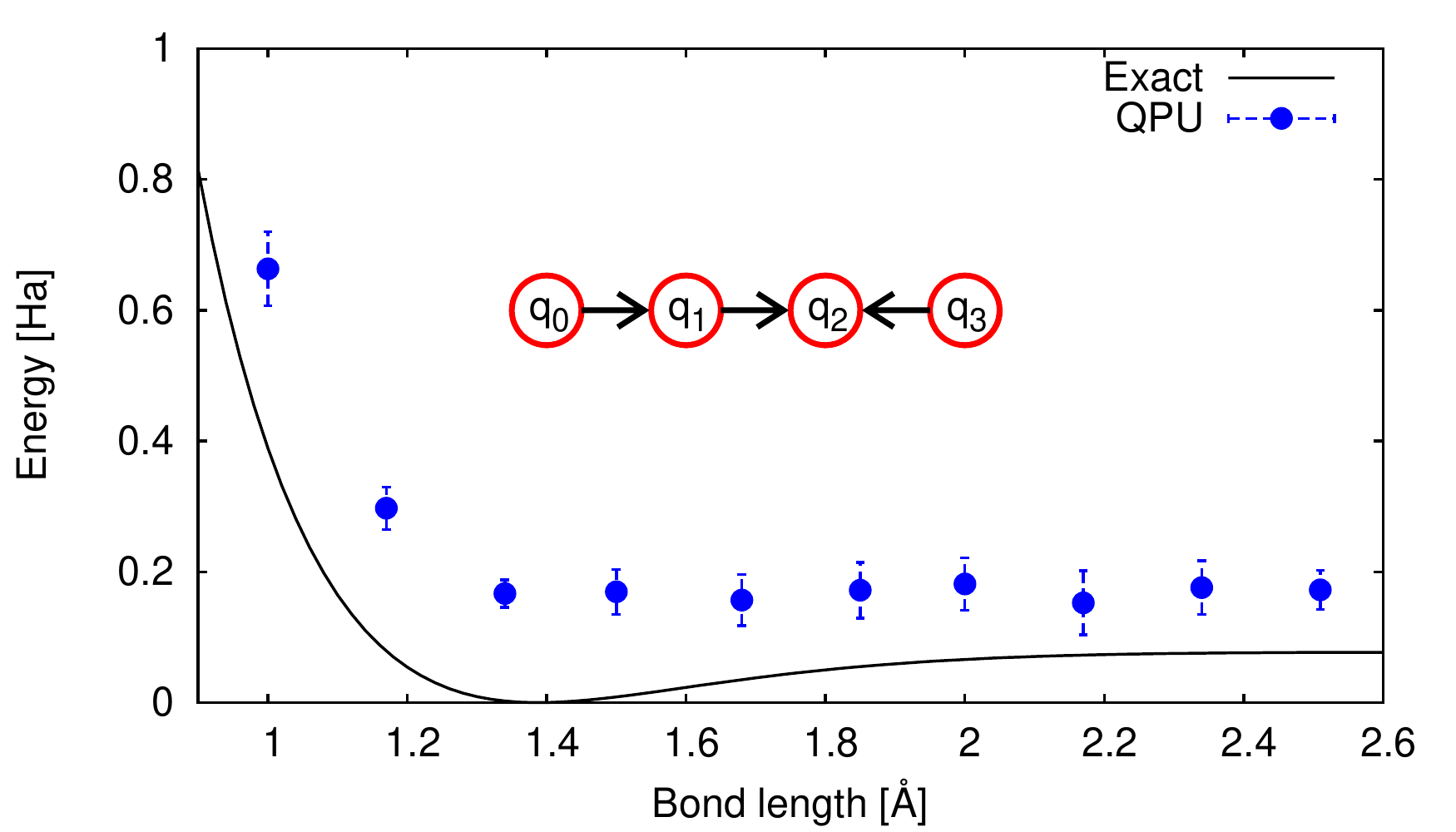}
   \caption{Calculated ground-state energy of F$_{2}$ as a function on bond length. The solid line gives the reference FCI result calculated with Psi4. The blue circles give the energies averaged over the last 25 iterations of the optimization, with the error bars representing the standard deviations of those sets. The results are from IBM's Poughkeepsie quantum computer. The inset shows the topology of the utilized qubits with the arrows indicating the employed CNOT gates. All energies are given with respect to the exact result at the equilibrium bond length.}
   \label{fig:pes}
\end{figure}
Figure~\ref{fig:pes} shows the results of our optimizations on the quantum computer. While the ideal simulation gave an average error of 1.07 kcal/mol across the potential energy curve, the actual device results give an average error of 90.9 kcal/mol. This is comparable to the average error of approximately 68 kcal/mol seen previously for the simulation of LiH on 4 qubits.\cite{Gambetta17_242} As we have already attempted to correct for readout errors, the remaining errors in the calculated energies relative to the ideal simulation could be attributable to noisy gate operations and decoherence/relaxation errors, which could potentially be mitigated with custom pulse sequences.\cite{Gambetta19_491}

In summary, we have described the SCM method, which provides compact system-to-qubits mappings saving scarce quantum computing resources. In particular, we demonstrated how this method can be used to achieve a factor of 4 reduction in the number of qubits needed, compared to standard mappings, to solve for the full many-particle ground state of the F$_{2}$ molecule in a minimal basis with 16 active spin-orbitals. We also showed how systems described by a real Hamiltonian benefit from a hardware-efficient ansatz that only includes single-qubit rotations around the $y$-axis, reducing the number of single-qubit operations roughly by a factor of three.

We conclude based on the above findings that the SCM method could be used to generate compact mappings to address systems several times larger than otherwise possible. If we assume that the development of quantum computing hardware follows a Moore's law with the quantum computing capability doubling every 3 years, say, a four-fold reduction in needed resources would correspond to 6 years of hardware development. As such gains are hard to pass up, we expect that SCM method will eventually become standard in simulating quantum mechanics on quantum computers, especially given that it can be used for bosonic as well as fermionic systems.

\subsection{Methods}
The FCI Hamiltonian in the basis of configuration states of the $\mathrm{A_g}$ irreducible representation of F$_{2}$ using the STO-3G atomic basis set\cite{Pople69_2657} was obtained from Psi4.\cite{Sherrill17_3185} This matrix was then decomposed into a sum of tensor products of Pauli operators to give the qubit representations (the qubit Hamiltonians are given in the SI). All of the quantum computer experiments were run via Qiskit,\cite{qiskit} and the trial state variational parameters were optimized via the simultaneous perturbation stochastic approximation algorithm with method parameters $a=1.2$, $A=20$, $c=0.06$, $\alpha=0.602$, and $\gamma=0.101$.\cite{Spall98_817} The 132 terms in the qubit Hamiltonians were grouped into 40 tensor product basis states via OpenFermion.\cite{openfermion} For the optimization, 200 iterations were performed and 1,024 shots were used for the energy evaluations at the perturbed parameters at each iteration. The readout error calibration matrix was calculated at the beginning of the optimization, and then every 10 iterations thereafter. We used 8,192 shots for the construction of the calibration matrix. During the last 25 iterations, the energies at the current parameters were also evaluated, again using 8,192 shots. The reported energies correspond to the average over these last 25 evaluations.

\subsection{Acknowledgements}
This work has been supported by the U.S. Naval Research Laboratory and its Naval Innovation Science and Engineering (NISE) programs. We acknowledge IBM Q resources through a collaboration with AFRL/RI and thank Steve Hellberg, Steve Policastro, and Sergio Tafur for discussions.

\end{document}